\documentclass[ste,twoside]{stefano}

\usepackage{amsmath}
\usepackage{amssymb}
\usepackage{graphics}
\usepackage{rotating}
\usepackage{cite}
\usepackage{color}

\def\makeheadbox{{%
\hbox to0pt{\vbox{\baselineskip=10dd\hrule\hbox
to\hsize{\vrule\kern3pt\vbox{\kern3pt \hbox{ {\sc hep-th/0401126}}
\hbox{ {\sc Phys. Rev. D. {\bf 69}, 034006-5 (2004)}
\hspace*{8.0cm} {\color{blue}{$\boldsymbol{\Sigma \delta
\Lambda}$}} }
\kern3pt}\hfil\kern3pt\vrule}\hrule}%
\hss}}}

\def\0{\mbox{\tiny $0$}}
\def\1{\mbox{\tiny $1$}}
\def\2{\mbox{\tiny $2$}}
\def\3{\mbox{\tiny $3$}}
\def\4{\mbox{\tiny $4$}}
\def\5{\mbox{\tiny $5$}}
\def\6{\mbox{\tiny $6$}}
\def\7{\mbox{\tiny $7$}}
\def\8{\mbox{\tiny $8$}}
\def\9{\mbox{\tiny $9$}}
\def\nr{\mbox{\tiny NR}}
\def\eff{\mbox{\tiny eff}}
\def\Y{\mbox{\tiny Yuk}}
\def\H{\mbox{\tiny Hul}}
\def\sch{\mbox{\tiny Sch}}
\def\di{\mbox{\tiny Dir}}
\def\dar{\mbox{\tiny Dar}}
\def\max{\mbox{\tiny max}}
\begin{document}
%

\title{\large AMPLIFICATION OF COUPLING FOR YUKAWA POTENTIALS}

\author{
Stefano De Leo\inst{1}
\and Pietro  Rotelli\inst{2} }

\institute{
Department of Applied Mathematics, State University of Campinas\\
PO Box 6065, SP 13083-970, Campinas, Brazil\\
{\em deleo@ime.unicamp.br}
\and
Department of Physics, INFN, University of Lecce\\
PO Box 193, 73100, Lecce, Italy\\
{\em rotelli@le.infn.it}
}


\date{Submitted: {\em August, 2003} - Revised: {\em November, 2003}}

\abstract{ It is well known that Yukawa potentials permit bound
states in the Schr\"odinger equation only if the ratio of
exchanged mass to bound mass is below a critical multiple of the
coupling constant. However arguments suggested by the Darwin term
imply a more complex situation. By studying numerically the Dirac
equation with a Yukawa potential we investigate this
"amplification" effect.}



\PACS{ {03.65.Ge} \and  {03.65.Pm} \and  {13.10.+q}{}}






\titlerunning{Amplification of coupling for Yukawa potentials}

\maketitle


\section*{I. INTRODUCTION}

\noindent

In recent years neutrino oscillation experiments have all but
proved that neutrino mass states exist~\cite{NEU}. These are
linear combinations of neutrino flavor states. The neutrinos are
subject only to weak interactions (neglecting gravitation). More
precisely, they participate only in weak vertices (loop diagrams
in which a neutrino couples to a charged lepton virtual pair,
$W^{\pm}$, would of course allow for higher order electromagnetic
interactions via the charged virtual particles).

Since neutrinos have mass and couple via exchanged intermediate
mass bosons to leptons one may legitimately  ask if a
neutrino-lepton bound state could exist. Exchange of $W^{\pm}$,
$Z^{\0}$ means the treatment of Yukawa potentials and the question
of when bound states exist for Yukawa potentials~\cite{FLU}. Such
weak bound states would provide very interesting effects in
superconductivity as they would constitute a "bosonic atom" which
remind us of one of the multiple roles played by  Cooper pairs in
standard theory.

 The question of the existence, or not, of a bound state
for a given potential seems a simple theoretical question with a
straightforward method for finding the answer. Given a potential
and the corresponding reduced mass wave equation one solves it,
normally numerically, and sees if bound states exist. Bound states
are characterized by normalized (localized) solutions and a
discrete energy spectrum below the free particle threshold $E<m
\Leftrightarrow E_{\nr}=E-m<0$, where $E$ is the relativistic
energy and $E_{\nr}$ the non-relativistic energy of the particle
with reduced mass $m$. Either the existence of normed states or of
a discrete spectrum suffices to identify a bound state regime. Of
course, nothing is quite so simple. For example there are limits
for the validity of the use of potentials. Furthermore  coupling
constants have the "annoying" tendency to run and hence, are all
but constants.

 Even within the realm of non-relativistic quantum mechanics
  potential theory,
we have surprising subtleties. A one-dimen\-sio\-nal square well
always yields a bound state. A three-dimensional spherical well
must be sufficiently "deep" or extended to allow a bound state
solution~\cite{CTDL}. Relativistic wave equations open up numerous
conundrums. Can a bound state energy exist below the onset of
negative energy free wave solutions (Klein paradox~\cite{KLE,CD})?
How does particle statistics and in particular the Pauli exclusion
principle contribute to or modify this. The use of field theory
will be mentioned in the conclusions of this work.

The weak interactions are characterized by the exchange of very
massive particles ($W^{\pm}$, $Z^{\0}$ exchange), parity violation
that constitute a complicated (but calculable) mix of attractive
and repulsive forces. For simplicity, one can consider only the
$Z^{\0}$ exchange. $Z^{\0}$ exchange allows for a potential
treatment since it is well represented by a single Feynman  graph
in momentum space from which a Yukawa potential can be derived by
Fourier transform. As a consequence of the V-A nature of weak
interactions, it can be shown that according to the total spin
state of the two fermion system the potential is either attractive
or repulsive. A similar attractive/repulsive alternative occurs
for the isospin state in {\em p}-{\em n} system from which the
deuteron singlet bound state emerges. The limitation for the
existence of an electron-neutrino bound state is set by the
enormous value of $\frac{\mu}{m}$ where $\mu(\approx 91
\mbox{Gev})$ is the $Z^{\0}$ mass~\cite{PPB} and $m$ the reduced
neutrino mass. As we have said, oscillation experiments are
consistent with the existence of neutrino mass states but these
have masses of less than a few electron-volts.

In this paper, we wish to investigate a small part of this
question. We concentrate our attention upon the Yukawa potential
and ask what are the limits upon the exchanged mass for two
fermion bound states to exist. The attractive Coulomb potential
yields infinite bound states solutions independent of the coupling
strength for both the Schr\"odinger and Dirac equations. A Yukawa
potential will on the contrary not yield a bound state unless the
coupling is sufficiently "strong". This is proven only numerically
since no analytic solution is known for the Yukawa potential.
Using the Schr\"odinger equation and
\begin{equation}
\label{eqy}  V^{\Y}(r) = - \, \lambda  \,  \, \frac{e^{- \mu
r}}{r}~,
\end{equation}
($\lambda
> 0$ for bound states)
it is known that the condition for the existence of the lowest
lying S-states is
\begin{equation}
\lambda \, \, >  \, 0.84 \, \, \frac{\mu}{m}~.
\end{equation}
An analytic approximation to this result can be found by using the
Hulth\'en potential~\cite{FLU,PAT} in the Schr\"odinger equation,
\begin{equation}
V^{\H}(r) = - \, \lambda  \, \frac{2\mu}{e^{2\mu r} - 1 }~.
\end{equation}
The numerator in this potential has been chosen not only because
of dimensional requirements but also to agree with Eq.(\ref{eqy})
up to the linear term in $\mu r$. The imposition of a normalized
wave function here requires
\begin{equation}
\lambda \, \, >  \, \frac{\mu}{m}~.
\end{equation}
The radial solution for the Hulth\'en ground state is
\[
R(r,l=0) = 2 \, \sqrt{m\lambda \, ( \, m^{\2}\lambda^{\2} -
\mu^{\2} \, ) \, }  \, \, \mbox{$\frac{\sinh(\mu r)}{\mu r}$}\, \,
e^{-m\lambda \,r} ~.
\]
 An equivalent potential (with
analytic solution) to the Hulth\'en for the Dirac equation is not
known. We must qualify this last statement. If one allows for
vector as well as a scalar potential, and one makes a very
particular choice then one can find analytic solutions with a
Hulth\'en scalar potential~\cite{CPL}. This technique derives from
some ingenious suggestions by Alhaidari~\cite{ALH} for solving the
Dirac-Morse problem.

\section*{II. DIRAC EQUATION AND YUKAWA POTENTIALS}

\noindent

The Dirac equation in the presence of a general spherical
potential $V(r)$ can be written as
\begin{equation}
\label{dirac} E \, \, \Psi(\boldsymbol{r}) \, = \left( \,
\begin{array}{cc} ~~V(r) + m~~  & - i \, \boldsymbol{\sigma} \cdot \nabla \\
- i \, \boldsymbol{\sigma} \cdot \nabla  &  ~~V(r) - m~~
\end{array} \right) \, \Psi(\boldsymbol{r})~.
\end{equation}
By using
\[
\label{psi} \Psi(\boldsymbol{r}) \, = \left(
\begin{array}{r} f_{j}^{k}(r) \, \mathcal{Y}_{jm_j}^k(\hat{\boldsymbol{r}}) \\
i \, g_{j}^{k}(r) \, \mathcal{Y}_{jm_j}^{-k}(\hat{\boldsymbol{r}})
\end{array}
\right)~,
\]
where $\mathcal{Y}_{jm_j}^k(\hat{\boldsymbol{r}})$ are the
spherical functions obtained by summing the spherical harmonics
$Y_{l,m_j\pm\frac{1}{2}}$ with the spinor eigenstates and $k= \pm
( \, j + \frac{1}{2} \,)$ for $l=j \pm \frac{1}{2}$, for details
see ref.~\cite{GRO93}, Eq.(\ref{dirac}) reduces (dropping the
subscripts and superscripts for the radial functions $f_{j}^{k}$
and $g_{j}^{k}$) to two coupled first order ordinary differential
equations
\begin{eqnarray}
\label{eqfg}
 \left[ \, E - m - V(r) \, \right] \, f(r)  & = &  -
\, g\,'(r) - \, \frac{1-k}{r} \, g(r)~, \nonumber \\
 & & \\
\left[ \, E + m - V(r) \, \right] \, g(r)  & = & f\,'(r) + \,
\frac{1+k}{r} \, f(r)~. \nonumber
\end{eqnarray}
It is instructive to make the non-relativistic limit in
Eqs.(\ref{eqfg}) by setting $E=m-\epsilon$ ($\epsilon >0$ for
bound states) and assuming $\epsilon \ll m$. Eliminating $g$ (the
"small" components) yields a second order equation for $f$ (we
drop both $\epsilon$ and $V(r)$ with respect to $2m$ in the second
of Eqs.(\ref{eqfg}))
\begin{equation}
\label{nrl} \left[ \, \epsilon + V(r) \, \right] \, f(r) =
\frac{f\,''(r)}{2m} + \frac{f\,'(r)}{mr} - \frac{k(k+1)}{2mr^{\2}}
\, f(r)
\end{equation}
or equivalently
\begin{equation}
- \frac{f\,''(r)}{2m} - \frac{f\,'(r)}{mr} + \left[ \,
\frac{l(l+1)}{2mr^{\2}} + V(r) \, \right] \, f(r) = - \, \epsilon
\,  f(r)~.
\end{equation}
This is just the radial part of the Schr\"odinger equation
\begin{equation}
\left[ \,  - \, \frac{\nabla^{^{\2}}}{2m} + V(r) \, \right] \,
\phi(\boldsymbol{r}) = E_{\nr} \, \phi(\boldsymbol{r})
\end{equation}
where $E_{\nr}\equiv - \epsilon <0$ and $\phi(\boldsymbol{r}) =
f(r) Y_{lm_l}(\hat{\boldsymbol{r}})$.

A more sophisticated limit exists, where relativistic corrections
are maintained (up to order $\boldsymbol{p}^{\4}$). This equation
can be derived either by a Foldy-Wouthuysen
transformation~\cite{FW}, or in the more heuristic method used by
Sakurai~\cite{SAK} in which the $f$ function is corrected in order
to be normalized to order $\boldsymbol{p}^{\4}$. This equation
reads (always assuming a spherical potential so that $\nabla V(r)
= V\,'(r) \, \hat{\boldsymbol{r}}$)
\begin{equation}
H_{\eff} \, \phi(\boldsymbol{r}) = E_{\nr} \, \phi
(\boldsymbol{r})
\end{equation}
with
\begin{equation}
\label{heff} H_{\eff} = \frac{\boldsymbol{p}^{\2}}{2m} + V(r) -
\frac{\boldsymbol{p}^{\4}}{8m^{\3}} + \frac{1}{8m^{\2}} \, \left[
\,  \nabla^{^{\2}} V(r) \, \right] + \frac{1}{4m^{\2}r} \, V\,'(r)
\, \boldsymbol{\sigma} \cdot \boldsymbol{L}~.
\end{equation}
Now consider the Yukawa potential $V^{\Y}(r)$. For the case $l=0$
(S-wave) Eq.(\ref{heff}) reduces to
\begin{equation}
H_{\eff}^{^{\mbox{ \tiny [$l$=$0]$}}} =
\frac{\boldsymbol{p}^{\2}}{2m} -
\frac{\boldsymbol{p}^{\4}}{8m^{\3}} - \lambda \, \left[ \, \left(
\, 1 + \frac{1}{8m^{\2}} \, \nabla^{^{\2}} \right) \, \frac{e^{-
\mu r}}{r} \, \right] ~.
\end{equation}
The last term in the above equation contains the Darwin delta term
 \begin{equation}
 \label{nabla}
\nabla^{^{\2}} \, \frac{e^{-\mu r}}{r} = \mu^{\2} \, \frac{e^{-\mu
r}}{r} -  4 \pi \, \delta^{\3}(\boldsymbol{r})~.
\end{equation}
This term is essential in the Coulomb case ($\mu=0$) to reproduce
to order $\boldsymbol{p}^{\4}$ the bound state energy dependence
on $n$ (principal quantum number) and $j$ only. When $\mu \neq 0$,
we note that this term contains an additional piece proportional
to $\mu^{\2}$ and hence with the same sign as the original
potential term. It produces an effective potential with the
coupling constant "amplified"
\begin{equation}
V_{\eff}(r) = - \, \lambda \, \left( \, 1 + \frac{ \mu^{\2}}{8 \,
m^{\2}} \, \right) \, \frac{e^{-\mu r}}{r}~.
\end{equation}
If this result is combined with the numerical Schr\"odinger
calculations mentioned previously we might expect the condition
for the existence of a bound state to be
\begin{equation}
\label{ly} \lambda_{\eff} = \lambda \, \left( \, 1 + \frac{
\mu^{\2}}{8 \, m^{\2}} \, \right) \, >  \, 0.84 \,  \,
\frac{\mu}{m}~.
\end{equation}
If true not only would there be bound states for a given $\lambda$
with values of $\frac{\mu}{m}$ higher than otherwise expected, but
far more spectacularly, for $\frac{\mu}{m}\gg 1$ the condition for
the existence of a bound state is "inverted" and reads
\begin{equation}
\lambda \, > \, 6.62 \,  \, \frac{m}{\mu}~~~~~~~~(\,
\mbox{$\frac{m}{\mu}$} \ll 1\,)~.
\end{equation}
Of course such a conclusion is highly speculative since it is
based upon the combination of Schr\"odinger results and only a
part of the relativistic correction to the Schr\"odinger equation.
Higher order corrections could greatly modify this hypothesis. At
this point it is logical  to go directly to the fully relativistic
Dirac equation and (numerically) solve it for the Yukawa
potential. We wish to see if there is an amplification effect and
its comparison with Eq.(\ref{ly}).

All our results are conveniently expressed in terms of the
dimensionless parameters
\[
\lambda~,~~~w = \frac{\mu}{m \, \lambda}~~~\mbox{and}~~~ \eta =
\left( \, \mbox{$\frac{E}{m}$} -1 \, \right) /
(\sqrt{1-\lambda^{\2}} - 1)~.
\]
Using these variables Eqs.(\ref{eqfg}) can be written as
\begin{eqnarray}
\label{eqfg2} v\,'(x) \, -  \, \frac{k}{x} \,  \, v(x) \, + \,
 \left[ \, \eta  \,  \mbox{$\frac{ \sqrt{1-\lambda^{^{\2}}} - 1}{
 \lambda}$}
   \, + \, \lambda \,
 \frac{e^{-wx}}{x} \, \right] \, u(x)  & = &  0~,\nonumber \\
 & & \\
u\,'(x) \, +  \, \frac{k}{x} \, \, u(x) \, - \,
 \left[ \, \frac{2}{\lambda} \, + \, \eta  \,
 \mbox{$\frac{ \sqrt{1-\lambda^{^{\2}}} - 1}{\lambda}$} \, +  \, \lambda \,
 \frac{e^{-wx}}{x} \, \right] \, v(x)  & = &  0~,\nonumber
\end{eqnarray}
where $x=m \, \lambda \, r$,  $u=f/x$ and $v=g/x$. Note that here
$\lambda$ appears explicitly. The Schr\"odinger limit ($\lambda
\ll 1$) is particular in that it can be expressed purely in terms
of $\eta$ and $w$ with no explicit dependence upon $\lambda$,
\begin{equation}
\label{usc} u\,''_{\sch}(x) \, - \, \frac{l(l+1)}{x^{\2}} \, \, \,
u_{\sch}(x) \, -  \, \eta \, u_{\sch}(x) \, + \, 2 \,
\frac{e^{-wx}}{x} \, u_{\sch}(x) = 0~~~~~~~~[\, u_{\sch}(x) \equiv
f(x) /x \, ]~.
\end{equation}

The asymptotic behavior for $x\to \infty$ and $x \to 0$ are
substantially different in the Schr\"odinger and Dirac equations.
In the former, we have
\[ u_{\sch}(x \to \infty) ~\to~ e^{-\sqrt{\eta} \, x}~,\]
while
\[ u_{\di}(x \to \infty) ~\to~
e^{-\frac{\sqrt{E^{^{\2}} - m^{\2}}}{m\,\lambda} \, \,
x}~~~\mbox{and}~~~v_{\di}(x \to \infty) ~\to~
e^{-\frac{\sqrt{E^{^{\2}} - m^{\2}}}{m\,\lambda} \, \, x}~.
\]
The Dirac limits constrain  the possible values of the bound state
energy by $E^{^{\2}} < m^{\2}$ or   $m>E>-m$, otherwise the
solution will not be localized. The lower limit is an example of
the Klein paradox, here involving  a bound state fermion. Indeed,
since $V(\infty)=0$ for $E<-m$ we would have at infinity $E-V<-m$,
which for a step potential is exactly the condition for the Klein
paradox .

There is also a substantial difference between the Schr\"odinger
and Dirac equations in the limit of $x\to 0$ (recall that $x$ is
proportional to the radial parameter). For Schr\'odinger one has
\[ u_{\sch}(x\to 0)~\to~ x^{\mbox{\tiny $l$+$1$}}~,\]
while
\[ u_{\di}(x \to 0) ~\to~
x^{\mbox{\tiny $\nu$+$1$}}~~~\mbox{and also}~~~v_{\di}(x \to 0)
~\to~ x^{\mbox{\tiny $\nu$+$1$}}~,
\]
with  $\nu^{\2} = k^{\2} - \lambda^{\2}$. Since $\nu$ must be real
and greater than $- \, \frac{1}{2}$ for a normalizable solution
this sets a limit upon the size of $\lambda$ whose value depends
upon $k$ i.e. angular momentum. Given that the minimum value of
$\nu^{\2}$ is zero, we have that
\begin{equation}
\lambda^{\2} \leq k^{\2}~.
\end{equation}
For the S-wave ($k=-1$) which we expect to be the lowest energy
bound state (if a bound state exists), we obtain
\begin{equation}
\lambda_{\mbox{\tiny max}}^{\mbox{\tiny S-wave}} \leq 1~.
\end{equation}
We also recall the well-known result~\cite{GRO93} that since
\[ \nu = \sqrt{1-\lambda^{^{\2}}} < 1~~~~~(\,k=-1\,)~.
\]
the Dirac S-wave function is infinite at $x\to 0$, both for Yukawa
and Coulomb potentials. This is in starch  contrast with the
Schr\"odinger results.

In Figs.\,1 and 2, we show the numerical results for the S-wave
ground state. In Fig.\,1 the Dirac equation result of $\eta$
versus $w$ is plotted. As the ratio of exchanged to bound state
mass $\frac{\mu}{m}$ is increased (increasing $w$) the value of
$\eta$ is reduced. This corresponds to the bound state energy
increasing towards the limit of $E=m$ beyond which the bound state
ceases to exist. For the case displayed with $\lambda=0.01$, we
essentially reproduce the Schr\"odinger equation curve, which is
$\lambda$ independent as long as $\lambda \ll 1$. This curve
confirms  the well known condition $\frac{\mu}{m} < 1.19 \,
\lambda$ or $w<1.19$ for Schr\"odinger. On the same graph, we
display the results of the numerical solutions to the Dirac
equation for various values of $\lambda$. As $\lambda$ increases
the Dirac results yield higher values of $w_{\max}$ and hence
evidence for an amplification effect. In Fig.\,2, we plot an
interpolation curve of $w_{\max}$ against $\lambda$ (up to its
maximum S-wave value of 1) for the Dirac results. The
Schr\"odinger curve is a flat line at 1.19 on this plot. Again the
magnification effect is evident. As a comparison, we display in
the same plot the prediction of the Darwin term, which stimulated
this analysis. For $\lambda<0.8$ the Dirac amplification is lower
than for the Darwin term, but it exceeds the latter for $\lambda
\sim 1$. At $\lambda=1$ the Schr\"odinger equation predicts as a
limit for an S-wave bound state
\begin{equation}
\left( \, \frac{\mu}{m} \, \right)_{\max}^{\sch} = 1.19~,
\end{equation}
with Dirac we find
\begin{equation}
\left( \, \frac{\mu}{m} \, \right)_{\max}^{\di} = 1.68~,
\end{equation}
while the Darwin term yields
\begin{equation}
\left( \, \frac{\mu}{m} \, \right)_{\max}^{\dar} = 1.54~.
\end{equation}

\section*{III. CONCLUSIONS}

Within the validity of the Dirac equation ($\lambda^{\2} \leq
k^{\2}$), we have confirmed by numerical calculations that the
effective Yukawa coupling constant is amplified with respect to
the non-relativistic Schr\"odinger potential. However, we have not
detected the spectacular "turn-over" phenomena as suggested by the
Darwin term for the Yukawa potential. At the very least our
results imply that bound states exist for higher mass exchanges
than otherwise expected. We have also identified in the analytic
studies of Gu, Zheng and Xu~\cite{CPL} the presence of an
amplification effect. However, the particular choice of potential
makes this result of doubtful application.

We are somewhat unsatisfied by the limits upon coupling constant
and bound state energy set by the asymptotic Dirac equations. To
go beyond this, we must necessarily use field theory. This
involves a numerical analysis of the Bethe-Salpeter
equation~\cite{BS}, and such an analysis is indeed planned.
Furthermore, the limitation set by the Klein effect ($E>-m$) is
very interesting for a bound state fermion. Conventionally, this
effect is interpreted by invoking pair production~\cite{CD}.
Since, an attractive (binding) potential for say a fermion is
repulsive for the corresponding anti-fermion we expect, if pair
production occurs, an anti-fermion flux to leak from the system
while the density of trapped fermions increases. However, the
Pauli exclusion principle will eventually block an increase in the
number of bound fermions. For example, with Yukawa there may be
only one bound state level which could accommodate at most two
spin $\frac{1}{2}$ fermions. This suggests that the Pauli
principle could block the Klein effect and allow for bound states
with $(E<-m$).


\newpage

\begin{figure}[hbp]
\hspace*{-2.5cm}
\includegraphics[width=10cm, height=19cm, angle=90]{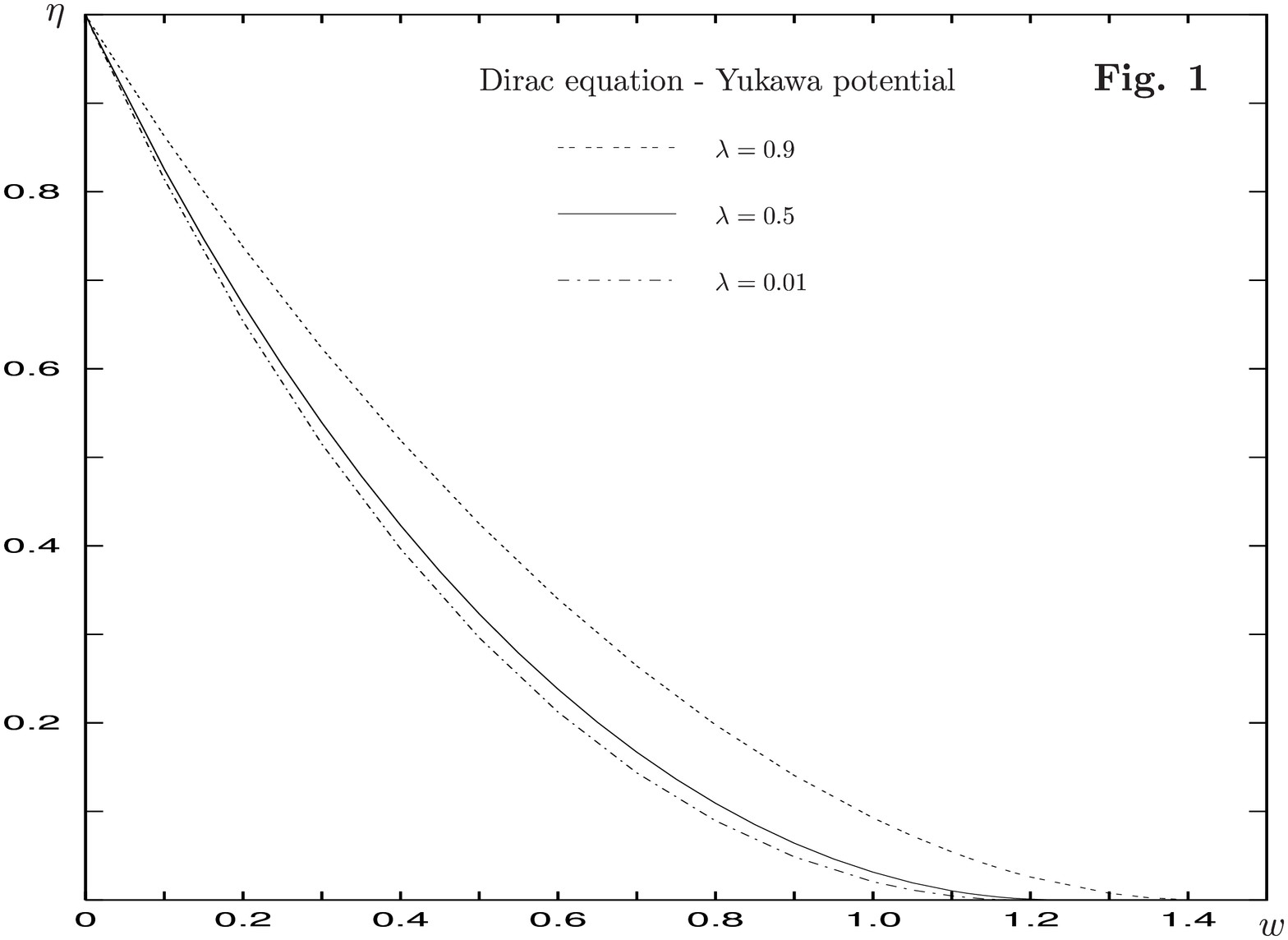}
\caption{$\eta$ versus $\omega$ for the ground state of the Yukawa
potential in the Dirac equation. Three values of $\lambda$ are
shown. For $\lambda \ll 1$ (the dot-dash curve) we reproduce
essentially the Schr\"odinger equation result. The value of
$w_{\max}$ when $\eta=0$ shows an increase with increasing
$\lambda$.} \hspace*{-2.5cm}
\includegraphics[width=10cm, height=19cm, angle=90]{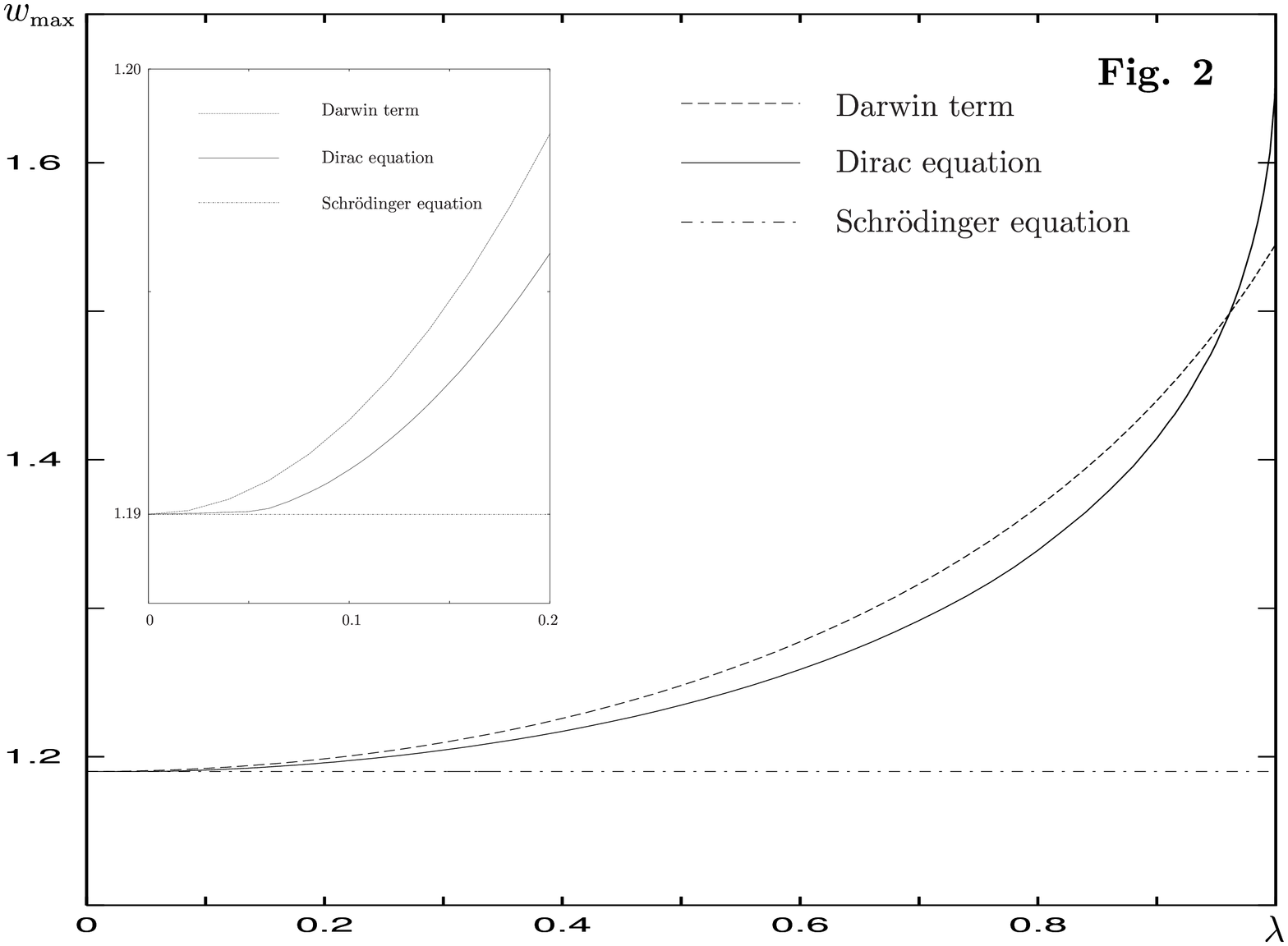}
\caption{ Comparison of $w_{\max}$ against $\lambda$ for the
ground state in the Dirac and Schr\"odinger equations. The dashed
curve is the magnification effect of the Yukawa version of the
Darwin term alone.}
\end{figure}

\end{document}